\documentclass[twocolumn]{aastex7}

\usepackage{amsmath}
\usepackage{subcaption}  
\definecolor{blue-violet}{rgb}{0.30, 0.1, 0.89}

\begin{document}

\title{High Energy Emission from the Intrabinary Shocks in Redback Pulsars}

\author[0000-0002-9545-7286]{Andrew G. Sullivan}
\affiliation{Kavli Institute for Particle Astrophysics and Cosmology, Department of Physics, Stanford University, Stanford, CA 94305, USA}
\email[show]{ags2198@stanford.edu} 

\author[0000-0001-6711-3286]{Roger W. Romani}
\affiliation{Kavli Institute for Particle Astrophysics and Cosmology, Department of Physics,
Stanford University, Stanford, CA 94305, USA}
\email{rwr@stanford.edu}



\begin{abstract}
The intrabinary shocks (IBS) of spider pulsars emit non-thermal synchrotron X-rays from accelerated electrons and positrons in the shocked pulsar wind, likely energized by magnetic reconnection. In redback spider pulsars, the IBS typically wraps around the pulsar, leading to a near-normal IBS shock with relatively bright X-ray emission. The characteristic energies of radiating particles and the magnetic fields in the IBS suggest spectral features in the hard X-ray band. Here we perform joint soft-hard X-ray analyses of three redback pulsars, J1723-2837, J2215+5135, and J2339-0533, including new J2215 NuSTAR data. We identify a significant cooling break in J1723-2837 and a marginal break in J2215+5135, while placing constraints on the break energy in J2339-0533. Interpreting these as synchrotron cooling features allows us to estimate the IBS magnetic field $B_{\rm IBS}  \sim 40-100$\,G and place lower bounds on the maximum radiating electron energy. Our results constrain the magnetization of the pulsar wind as well as pair-production in millisecond pulsar magnetospheres.
\end{abstract}



\section{Introduction} \label{sec:intro}
Redback pulsars are millisecond pulsars (MSPs) in $P_b\lesssim1$ day orbits with $M_c\approx 0.1-0.4$ M$_\odot$ stellar companions. The irradiation from pulsar gamma-rays and wind relativistic particles drives a companion outflow \citep{1988Natur.334..225K,1988Natur.334..684V, 2013IAUS..291..127R, 2019Galax...7...93H} which collides with the relativistic pulsar wind to form an intrabinary shock (IBS). For redbacks, this shock generally wraps around the pulsar \citep{2016ApJ...828....7R, 2017ApJ...839...80W, 2019ApJ...879...73K}, producing shocks nearly normal to the incident striped wind over a large area and generating relatively bright X-ray emission.

Most redbacks have X-ray orbital light curves with one or two peaks \citep[e.g.][]{2014ApJ...781....6B, 2019ApJ...879...73K, 2021ApJ...917L..13K, 2024ApJ...974..315S}. These are due to beamed synchrotron emission from radiating particles traveling parallel to the contact discontinuity in the shocked pulsar wind \citep[e.g.][]{2019ApJ...879...73K,2022ApJ...933..140C}. The non-thermal IBS spectra show extremely hard power-laws $dN_\gamma/dE\propto E ^{-\Gamma}$, often with $\Gamma\lesssim1$ \citep[e.g.][]{2019ApJ...879...73K}. Such hard spectra suggest shock-induced magnetic reconnection {could be} the main particle acceleration mechanism \citep{2011ApJ...741...39S, 2022ApJ...933..140C, 2024MNRAS.534.2551C}. 

At the shock, the incompletely annihilated striped wind magnetic field can result in synchrotron cooling timescales comparable with the flow time along the IBS, which can induce spectral cooling breaks \citep{2019ApJ...879...73K, 2024ApJ...964..109S}. Furthermore, exponential cutoffs in the synchrotron spectrum correspond to the maximum energy of the electron and positron ($e^\pm$) population, which, {for reconnection,} depends on the magnetization of the pre-shock pulsar wind \citep{2024MNRAS.534.2551C} and, by extension, the magnetospheric pair multiplicity \citep{2022ARA&A..60..495P}. Thus, high energy spectral features probe the pulsar wind physics.

Here we study X-ray emission from redbacks J1723-2837 (J1723), J2215+5135 (J2215), and J2339-0533 (J2339), using archival  \textit{XMM}-Newton and NuSTAR observations of J1723 and J2339 and new NuSTAR data for J2215 to measure the spectral indices and constrain hard X-ray spectral features. Sec. \ref{sec:Observation} summarizes the observations and presents \textit{XMM}-Newton and NuSTAR light curves for these sources. Sec. \ref{sec:Spectra} describes analyses of the phase-dependent X-ray spectra. We discuss our results and their implications in Sec. \ref{sec:discussion}.

\section{Observational Data}
\begin{figure*}

\hspace*{-2.2cm}
    \includegraphics[width=1.25\linewidth]{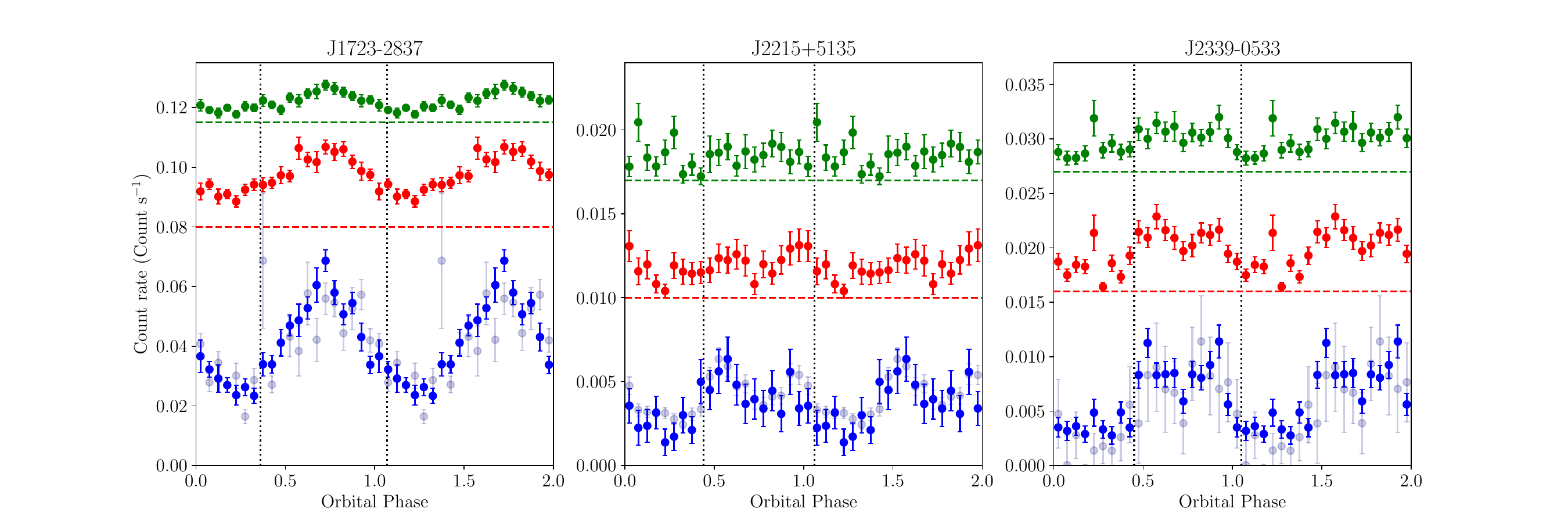}
    \caption{NuSTAR light curves for the three redbacks in different energy ranges. Bottom is $3-10$ keV, middle is $10-20$ keV, and top is $20-40$ keV. The faded markers overlaid with the NuSTAR $3-10$ keV show the XMM light curves in the same energy range. The dashed lines show the 0 count s$^{-1}$ level for each light curve.}
    \label{fig:lightcurves}
\end{figure*}
\label{sec:Observation}
\subsection{J1723-2837}
J1723 is a radio pulsar with spin period $P=1.86$ ms in a $P_B=14.8$ hr orbit with a $0.4-0.7$ M$_\odot$ companion \citep{2013ApJ...776...20C}. X-ray analyses of the soft \citep{2014ApJ...781....6B} and hard X-rays \citep{2017ApJ...839..130K} suggest that the emission is IBS dominated, with a hard spectrum peak in the light curve. Here, we jointly analyze the two data sets for broad-band spectral features.

We analyze the 56.3\,ks \textit{XMM} observation (ObsID 0653830101; \cite{2014ApJ...781....6B}) with the XMM-Newton Scientific Analysis System (SAS) \citep{2001A&A...365L...1J}.  The data were processed using the \texttt{epproc} and \texttt{emproc} commands and cleaned of background flares, leaving 28.7\,ks with the EPIC-PN and 31.8\,ks with the EPIC-MOS cameras. We extract the source from a 20" radius circular aperture and use background regions 6x and 25x larger for PN and MOS, respectively. For the 168\,ks NuSTAR exposure (ObsID 30101043002; \cite{2017ApJ...839..130K}) we used the standard \texttt{nupipeline} script to extract the source from a 38" radius circular aperture, with a 86" radius circular background region to generate the 3-78 keV light curves and spectra, applying standard exposure and barycenter corrections. We show the NuSTAR and XMM light curves in the left panel of Fig.\,\ref{fig:lightcurves}.

\subsection{J2215+5135}
J2215 is a $P=2.6$\,ms Fermi-LAT gamma-ray pulsar in a $P_B=4.14$\,hr orbit with a $0.3\,M_\odot$ companion \citep{2023ApJ...958..191S}. Recent 0.5-10\,keV XMM-Newton analyses \citep{2024ApJ...974..315S} show IBS emission dominating the X-rays. Two well-resolved, asymmetric IBS peaks widely separated in orbital phase indicate a rather flat IBS wrapped around the pulsar and distorted by companion orbital motion.

We report NuSTAR observations of J2215, which took place 2024, March 4-6 for 194\,ks (ObsID: 30901014002). We reduced the data using the standard \texttt{nudas} module in \texttt{HEADAS}, extracting the source 3-78\,keV light curves and spectra from a 32" radius circular aperture with an 80" radius circular background using \texttt{nupipeline} and applying standard exposure and barycenter corrections. We supplement these NuSTAR observations with $0.5-10$\,keV XMM Newton data (ObsIDs 0783530301, 0900770101, 0900770201, 0900770301). In the middle panel of Fig.\,\ref{fig:lightcurves}, we show the NuSTAR 3-10\,keV, 10-20\,keV, and 20-40\,keV light curves. The source is not well detected above $\sim 40$\,keV.

\subsection{J2339-0533}
J2339, also a gamma-ray pulsar, has spin period $P=2.88$\,ms and binary orbital period $P_B=4.6$\,hr with a 0.3\,M$_\odot$ companion \citep{2015ApJ...807...18P}.
The available archival XMM (182 ks, ObsIDs 0721130101, 0790800101) and NuSTAR (164\,ks ObsID 30202020002) observations have already been jointly analyzed \citep{2019ApJ...879...73K}; we re-analyze them here for consistency. Standard SAS flare cleaning and reduction leave 100.2\,ks from PN, and 115.6\,ks from MOS. The NuSTAR data were reduced using standard \texttt{nudas} methods. The right panel of Fig.\,\ref{fig:lightcurves} shows the J2339 light curves. Note that the XMM errors are larger than the other sources; we do not include PN in showing the light curve as it was in timing mode rather than imaging mode.

\section{Spectral Analysis}
\label{sec:Spectra}
Previous analyses show substantial orbital phase modulation of redback spectra, {with significant differences in hardness between flux maxima and minima \citep{2014ApJ...781....6B, 2019ApJ...879...73K, 2024ApJ...974..315S}}.
We therefore divide our data into two orbital phase intervals: an IBS region containing the 0.5-10\,keV XMM light curve peaks and an Off region with the flux minimum. The Off region may include softer non-IBS fluxes (e.g.\,from the neutron star, magnetosphere or pulsar wind nebula), as only weakly beamed X-rays from the IBS nose are expected; {however, our analysis is agnostic to the source of these fluxes. } In Table \ref{tab:Phasetable}, we list the phase intervals chosen for the three sources. {We find that altering the phase region boundaries or centers by $\Delta\Phi \sim 0.1$ alters the derived power-law spectral indices by $<1\sigma$.}
\begin{table}[]
    \centering
    \setlength{\tabcolsep}{4pt}
    \begin{tabular}{ccc}
     \hline\hline
    Object & IBS Phase & Off Phase\\
    \hline         J1723-2837& $0.36<\Phi<1.07$ & $0.07<\Phi<0.36$\\
        J2215+5135 & $0.44<\Phi<1.07$ & $0.07<\Phi<0.44$\\

         J2339-0533 &$0.45<\Phi<1.05$ & $0.05<\Phi<0.45$ \\
        \hline\hline
    \end{tabular}
    \caption{IBS and Off phase Intervals}
    \label{tab:Phasetable}
\end{table}

We use the CIAO tool \texttt{SHERPA} \citep{2001SPIE.4477...76F, 2007ASPC..376..543D} to perform spectral fits. 
First, for each of the three sources, we simultaneously fit absorbed power-law spectral models to the XMM {and NuSTAR data}, {with a phase-independent $\Gamma_0$ and an additional $\Gamma_{IBS}$ component in the peak phase}.
The spectral model in counts s$^{-1}$ keV$^{-1}$ is therefore
\begin{equation}
\label{eq:specmodel}
    f(E)=\begin{cases} e^{-N_H \sigma(E)}\left(K_{IBS}E^{-\Gamma_{IBS}}+K_{0}E^{-\Gamma_0}\right) & \text{IBS}\\
    e^{-N_H \sigma(E)}K_{0}E^{-\Gamma_0} & \text{Off}
    \end{cases},
\end{equation}
where $N_H$ is the equivalent hydrogen column, $\sigma(E)$ is the photo-electric cross section, and $K_{IBS}$ and $K_{0}$ are the power-law normalization factors. For the absorption, we employ the photoelectric cross sections of \cite{1992ApJ...400..699B} and the elemental abundances of \cite{1989GeCoA..53..197A}. As the instrument responses are imperfectly known, we allow for different normalizations of the XMM and NuSTAR power-laws with common $\Gamma$. 

For each object, we compare the unbroken IBS power-law models to a set of models with high energy spectral features: 1) a ``cooled" IBS power-law, which has a $\Delta \Gamma_{IBS}=+0.5$ spectral break at energy $E_{b, IBS}$, as expected for a population of synchrotron cooling electrons, 2) an exponential cutoff in the IBS power-law, with the IBS term in eq. \ref{eq:specmodel} scaled by $\exp(-E/E_{c, IBS})$, and 3) a ``cooled" and exponentially  cutoff IBS power-law. For J1723, we additionally fit with cooling breaks in both the IBS and Off components, adding $E_{b,0}$ as a parameter. This phase-independent break model is highly disfavored for J2215 and J2339, so we do not report those results. {In principle, subtle changes in spectral index can originate from different injection spectra along the shock \citep{2017ApJ...839...80W, 2020ApJ...904...91V}; however, this injection tends to harden the spectrum. As we look specifically for softening, we are not probing such models.}

\begin{figure*}

    \includegraphics[width=\linewidth]{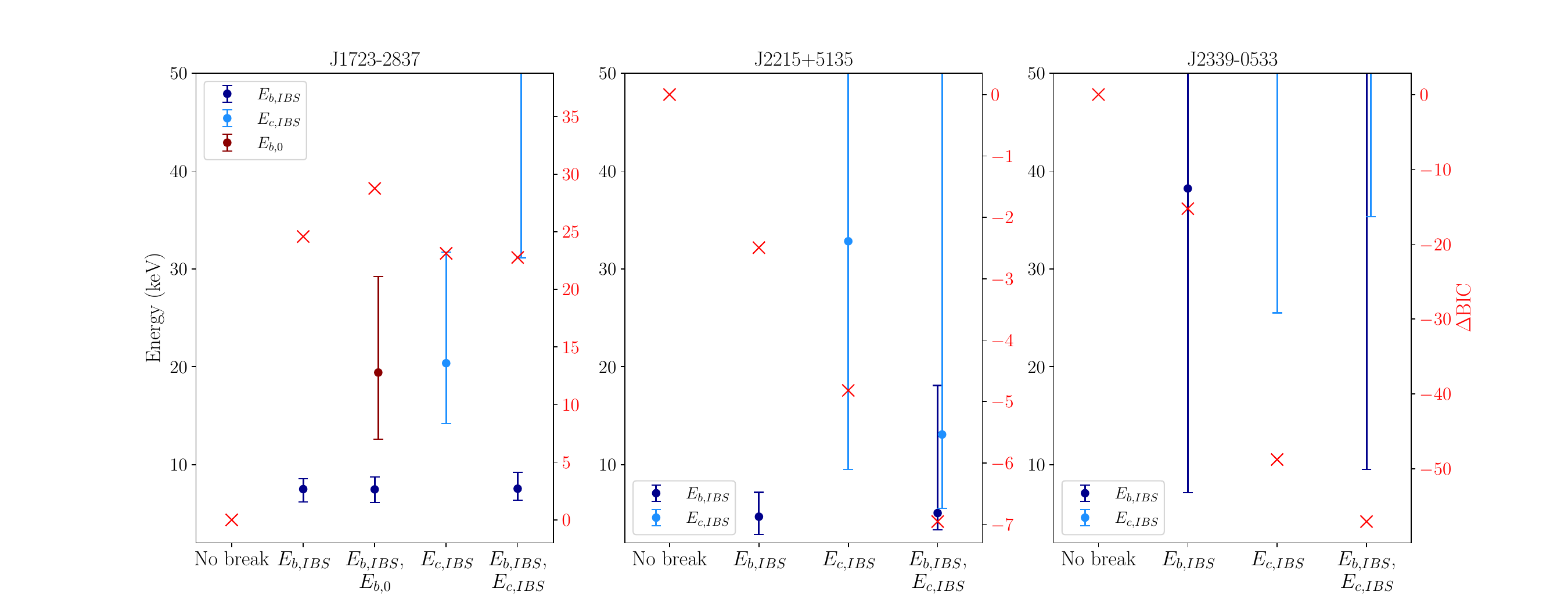}
    \caption{A comparison of the best-fitting spectral models to the unbroken power-law and constraints on estimated parameters for the three sources. The left axes plot the energy of the spectral feature (i.e. a cooling break or exponential cutoff). Error flags show $90\%$ confidence intervals. The right axes plot the $\Delta$BIC from the model without a break with crosses as the markers. }
\label{fig:redbackbreaks}
\end{figure*}

Our model fit results with 90\% confidence intervals for J1723, J2215, and J2339 are shown in Table \ref{table:2}. The first row for each object shows the base power-law model, while the subsequent rows show the models with spectral features. {The bold rows demarcate the BIC-preferred model for each source.} In Fig.\,\ref{fig:redbackbreaks}, we compare each source's different spectral models. We show the $\Delta$BIC from the unbroken power-law model as well as the 90\% confidence intervals on the estimated $E_b$ and $E_c$ values. Note that the crosses mark $\Delta$BIC from the simple power-law; high values are preferred.  In the cases of J2215 and J2339, the unbroken power-law model remains the best. {Intriguingly, a $E_{b, IBS}\approx5$\,keV break is nominally seen for J2215. However the $\chi^2$ decrease is small and BIC slightly decreases from the base model (see right hand scale, Fig.\,\ref{fig:redbackbreaks} center), so we do not select this model.} For J2339, we cannot make secure measurements of any breaks or cutoffs in the spectra and instead only report lower limits.

\begin{figure}
   \centering
    \includegraphics[width=\columnwidth]{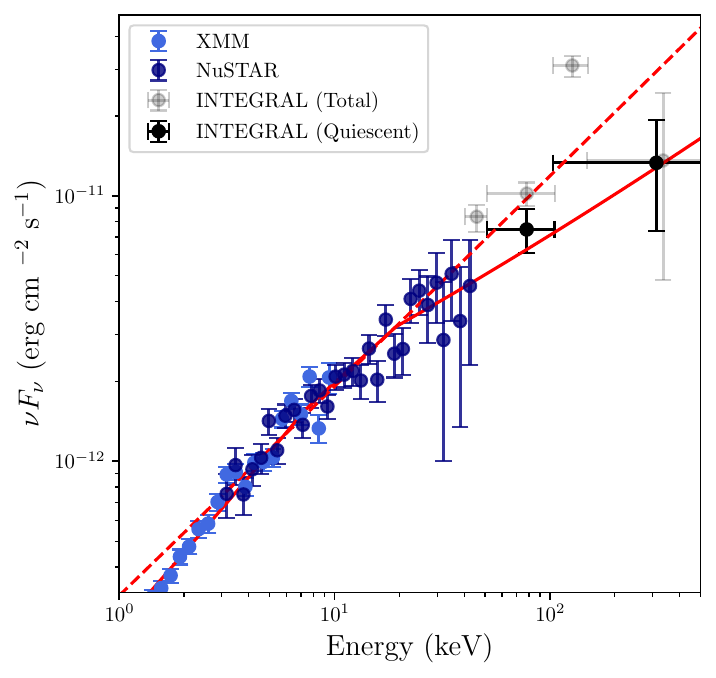}
    \caption{The XMM+NuSTAR spectra in the IBS region along with quoted fluxes from the INTEGRAL ISGRI catalog for J1723. The faded points show the flux in the total INTEGRAL exposure which are dominated by flares, while the dark points show flux in a quiescent 5.2 Ms exposure.  The solid line shows the best fitting cooled IBS and cooled Off power-law model, while the dashed line shows the best fitting unbroken power-law model.}
    \label{fig:J1723specInt}
\end{figure}

We detect a significant spectral break in J1723, with a $\Delta$BIC $>30$ from the unbroken model. In the left panel of Fig.\,\ref{fig:redbackbreaks}, we show models with only the IBS component cooled, a model with both IBS and Off power-laws cooled, an IBS power-law with an exponential cutoff, and an IBS power-law with both cooling and an exponential cutoff. All of these models show significantly improved $\Delta$BIC from the unbroken power-law. The model with the best $\Delta$BIC has cooled IBS and Off power-laws with $E_{b, IBS}=7.5_{-1.3}^{+1.3}$\,keV and $E_{b, 0}=19_{-7}^{+10}$\,keV and corresponds to $\dagger$ in Table \ref{table:2}. {Although the joint XMM and NuSTAR data sets are required to significantly detect it, the break is robust. The NuSTAR normalizations of the $\Gamma_0$ and $\Gamma_{IBS}$ broken power-laws differ by less than $1\sigma$ from the XMM power-law normalizations. Additionally, when fit independently, the unbroken XMM and NuSTAR spectra in the overlapping 3-10 keV band give consistent parameters.}

A model with just an exponential cutoff, while deprecated compared to those with cooling breaks, gives cutoff energy of {$E_{c, IBS}=20_{-6}^{+11}$}\,keV. Including both a cooling break and an exponential cutoff in the same IBS power-law gives {$E_{c, IBS}>32$\,keV. These models prefer a slightly harder $\Gamma_{IBS}=0.68_{-0.16}^{+0.15}$ and $\Gamma_{IBS}=0.79_{-0.18}^{+0.13}$, respectively.} In the model with only an exponential cutoff, the lower cutoff energy is required to absorb the features of the lower energy cooling break.

J1723 is present in the INTEGRAL-ISGRI source catalog with a 9.8 Ms exposure. The ISGRI catalog data show significant counts at $>100$ keV. Examining its light curve, short bright flares dominate the mission average flux, leading to a spectrum well above any reasonable extrapolation of our (apparently quiescent) observations. The high $\sim 130$\,keV catalog point in particular is anomalous. However, using the online INTEGRAL catalog tools, we extract from the interval MJD 52700-55200 where few bright flares are present. This corresponds to a 5.2\,Ms exposure. The flux from this interval has only two significant energy bins, but these are well consistent with our lower energy measurements and agree best with the cooled IBS and Off power-law model. The ISGRI points both with and without flares are shown along with the lower energy data and two model fits in Fig.\,\ref{fig:J1723specInt}. These data should raise the lower bound on $E_{c,IBS}$. A full ISGRI analysis of the flare and quiescent spectra of this system lies outside of the scope of this paper. As X-ray/optical flares are seen in other spiders \citep[e.g.][]{2015ApJ...804..115R, 2017ApJ...850..100A}, this could be of interest for future work.

\subsection{Comparison to Previous Analyses}
{We obtain $N_H$, $\Gamma_{IBS}$, and $\Gamma_0$ values for J1723 compatible with those of \cite{2014ApJ...781....6B}. Our phase-resolved analysis differs from \cite{2014ApJ...781....6B}, since they only analyze Chandra X-ray Observatory (\textit{CXO}) data and do not assume the component at flux minima is present at all phases.  The spectral indices for our BIC-preferred cooled IBS and Off model, however, are consistent with \citet{2014ApJ...781....6B}, as our spectral breaks are above the CXO band.} 
{Our best-fitting results for J2215 are consistent with \citet{2024ApJ...974..315S}, since additional spectral features are not required. We now place tighter constraints on the IBS and Off power-law indices.}

For J2339, we measure a low $N_H$. This lack of absorption means that very soft components can affect the fit to the power-laws, especially $\Gamma_0$. \citet{2019ApJ...879...73K} include additional data sets with soft response from {\it CXO} and {\it Suzaku}, along with a thermal neutron star polar cap emission component, in their model fitting. We choose not to include a thermal component.  This results in a harder estimated $\Gamma_0$, although $\Gamma_{IBS}$ remains consistent \citep{2019ApJ...879...73K}.

\begin{table*}
\setlength{\tabcolsep}{2pt}
\begin{tabular}{lccccccccc}
 
 \hline\hline
 Object & $B_{LC}$($10^5$ G) & $N_H$($10^{21}$ cm$^{-2}$) &  $\Gamma_{IBS}$ & $\Gamma_{0}$ & $E_{b, IBS}$ (keV)& $E_{c, IBS}$ (keV) & $B_{IBS}$ (G) &$\gamma_{max}$ ($10^5$) & $\chi^2/$DoF \\
 \hline
 J1723$-$2837 & 1.6 & 
$2.1^{+0.2}_{-0.2}$ & $1.04_ {-0.11}^{+0.11}$  &$1.22_{-0.06}^{+0.06}$ & & & & & $1.03$\\

 & & $1.9_{-0.2}^{+0.2}$ & $0.86_{-0.11}^{+0.11}$ & $1.23_{-0.06}^{+0.06}$ & 
$7.5_{-1.2}^{+1.2}$ & & $38^{+3}_{-2}$ & & 1.01\\
 &&$\mathbf{1.9_{-0.2}^{+0.2}}$ & $\mathbf{0.85_{-0.12}^{+0.13}}$ & $\mathbf{1.21^{+0.06}_{-0.06}}$ & $\mathbf{7.5_{-1.3}^{+1.3}}$$^\dagger$ &  & $\mathbf{38^{+2}_{-2}}$ &  & $\mathbf{1.01}$\\
 & & $2.0_{-0.2}^{+0.2}$ & $0.68_{-0.16}^{+0.15}$  & $1.23_{-0.06}^{+0.06}$ &  & $20_{-6}^{+11}$ & & $>1.7$& 1.01\\
  & & $1.9_{-0.2}^{+0.2}$ & $0.79_{-0.18}^{+0.13}$ & $1.23_{-0.06}^{+0.06}$ & $7.5_{-1.2}^{+1.6}$ & $>32$ & $38^{+2}_{-2}$& $>2.2$ & 1.01\\
\hline
J2215+5135 & 1.2&$\mathbf{0.38^{+0.38}_{-0.36}}$& $\mathbf{0.84_{-0.12}^{+0.13}}$ & $\mathbf{1.29_{-0.15}^{+0.16}}$ &  &  & &  & $\mathbf{1.08}$\\
 & &$<0.58$& $0.72_{-0.19}^{+0.15}$ & $1.26_{-0.15}^{+0.17}$ &  $4.7_{-1.8}^{+2.5}$ &  & $72_{-10}^{+13}$ &  & 1.07\\
    & &$<0.68$& $0.72_{-0.28}^{+0.24}$  & $1.27_{-0.16}^{+0.17}$ &   & $>9.5$ &  & $>1.0$  & 1.08\\
    & &$<0.52$& $0.50_{-0.19}^{+0.15}$ & $1.26_{-0.15}^{+0.17}$ &  $5.1_{-1.7}^{+13.0}$ & $>5.5$ & $71_{-25}^{+10}$ & $>0.6$ & 1.07\\ \hline
J2339$-$0533 & 0.73&$\mathbf{<0.38}$ & $\mathbf{0.94_{-0.20}^{+0.18}}$ & $\mathbf{1.71^{+0.42}_{-0.15}}$ &  &  &  & & $\mathbf{1.11}$ \\
 & &$<0.54$ & $0.82_{-0.40}^{+0.32}$ & $1.66_{-0.12}^{+0.56}$ & $>7.1$ &  & $<63$&  & 1.11 \\
 & &$<1.3$ & $1.06_{-0.56}^{+0.37}$ & $1.33^{+1.26}_{-0.03}$ &  & $>26$ & &$>1.9$ & 1.12 \\
  & &$<1.3$ & $1.09_{-0.63}^{+0.34}$ & $1.33^{+1.26}_{-0.03}$ & $>9.5$ & $>35$ & $<57$&$>1.9$ & 1.12 \\
 \hline\hline
\end{tabular}
\caption{High energy spectral models for three redback pulsars. 90\% confidence intervals on each parameter are reported. While only J1723 has a strong detection of spectral features, we place constraints on $E_c$ and $\gamma_{\rm max}$ for all pulsars. The BIC-preferred models for each pulsar are in bold. ($\dagger$)The preferred model for J1723 contains a cooled Off power-law with $E_{b,0}=19_{-7}^{+10}$ keV.}
\label{table:2}
\end{table*}

\section{Discussion and Conclusion}
\label{sec:discussion}
Measuring cooling breaks gives the magnetic field amplitude in the IBS if the flow time of the particles is known.  We estimate the flow time as
\begin{equation}
    t_f=\frac{a}{c\Gamma_{\rm bulk}\sqrt{1-\Gamma_{\rm bulk}^{-2}}}\approx 3 \text{ s}\left(\frac{a}{10^{11} \text{ cm}}\right)\frac{1}{\Gamma_{\rm bulk}\sqrt{1-\Gamma_{\rm bulk}^{-2}}},
\end{equation}
where $a$ is the binary orbital separation and $c$ is the speed of light \citep{2023IBSPolarization}. 
The cooling time of synchrotron radiating electrons is 
\begin{equation}
    t_c = \frac{3m_e^3 c^5}{2e^4\gamma B_{\perp}^2}\approx 50 \text{ s} \left(\frac{\gamma}{10^5}\right)^{-1}\left(\frac{B_{\perp}}{10 \text{ G}}\right)^{-2},
\end{equation}
where $\gamma$ is the Lorentz factor of the emitting particles and $B_\perp$ is the magnetic field perpendicular to the particle motion. Setting $t_c\approx t_f$ gives the particle Lorentz factor that corresponds to a cooling break $\gamma_b$ in terms of $B_\perp$. The cooling break energy is  
\begin{equation}
    E_b=\frac{3\hbar \gamma_b^2 e B_{\perp}}{2m_e c}\approx 10 \text{ keV} \left(\frac{\gamma_b}{10^5}\right)^2\left(\frac{B_{\perp}}{10 \text{ G}}\right).
\end{equation}
Thus, the magnetic field in the IBS is
\begin{equation}
    B_{IBS, \perp}=\frac{3}{2} \left ( \frac{m_e^5c^9\hbar}{e^7E_b t_f^2} \right )^{\frac{1}{3}}\approx65\,{\rm G}\, \left(\frac{E_b}{10\,{\rm keV}}\right)^{-\frac{1}{3}} \left(\frac{t_f}{3\,{\rm s}}\right)^{-\frac{2}{3}}.
\end{equation}
Scaling $B_{IBS, \perp}$ by $4/\pi$ gives the angle averaged $B_{IBS}$.

\cite{2024ApJ...974..315S} fit for the geometry and flow properties of the IBS of J2215 using the XMM X-ray light curves. The \texttt{ICARUS} IBS models compute the emitted X-ray flux as well as the local bulk flow speed at every position along the shock. To get a characteristic flow time, we compute the IBS phase flux weighted-average bulk Lorentz factor (at the best-fit J2215 viewing angle) as $\Gamma_{bulk}=1.4$. The geometry dependence (from different viewing angles and companion wind/pulsar wind momenta ratios) is weak, affecting $\Gamma_{bulk}$ by less than $<10\%$. Since the other redbacks should have similar geometry, we simply assume $\Gamma_{bulk}=1.4$ and scale with semi-major axis $a$ to get a characteristic $t_f$ for the other two sources. With upper limits on $E_{b, IBS}$ for J2339, we only provide an upper bound on the magnetic field, while we are able to estimate 90\% confidence ranges for J1723 and J2215. These characteristic IBS fields are in the eigth column of Table \ref{table:2}. 

With a toroidal pulsar wind magnetic field $B_{\rm wind} \sim B_{LC} (r_{\rm LC}/r)$, the pre-shock field should be $B\sim20-30$\,G for these three redbacks, so the cooling breaks provide plausible estimates for the compressed, but partly annihilated, post-shock magnetic field of the striped wind. Note that striped wind field annihilation should be more nearly complete at the pulsar spin equator \citep[e.g.\,][]{1999A&A...349.1017B, 2015MNRAS.448..606C}, assumed here to lie in the orbital plane. Thus it is interesting that J2215 and J2339 are apparently viewed at rather high inclinations $i \sim 64-80^\circ$  \citep[J2215,][]{2024ApJ...974..315S} and $i \sim 70^\circ$ \citep[J2339,][]{2020ApJ...903...39K}, so that the post-shock field in view should be largely annihilated. In contrast, J1723 at an estimated $i \sim 41^\circ$ \citep{2016ApJ...833L..12V} should preserve a large residual post-shock field. These $i$ estimates are supported by the IBS light curves, as both J2215 and J2339 have double-peaked {\it XMM} light curves, indicating a view through the IBS cone at high inclination, while J1723's apparent single peak indicates a grazing view of the cone at lower inclination. This provides a plausible explanation for why J1723, with its expected strong residual field, prefers a cooling break while the other sources do not. Our break sensitivity is best for J1723 as the brightest source as well.

Using the constraints on the magnetic field and exponential cutoff, we can put lower limits on the maximum energy of the IBS $e^\pm$ population. $E_{c,IBS}\gtrsim6$ keV in each source implies a maximum radiating particle energy $\gamma_{max}\gtrsim6\times10^4$ (assuming $E_{b,IBS}\gtrsim E_{c,IBS}$ in models where a cooling break is not fit), well below the radiation reaction limit $\gamma_{rad}$ \citep{1992ApJ...396..161D} and {Hillas limit $\gamma_H$ \citep{1984ARA&A..22..425H}} for the estimated $B_{IBS}$. The maximum reconnection-driven particle energy probes the pulsar wind magnetization $\sigma_{pw}={B^2}/{4\pi n_e m_e c^2}$, since reconnection efficiently accelerates particles to $\gamma_{max}\sim\sigma_{pw}$ \citep{2011ApJ...741...39S, 2022ApJ...933..140C}. Note that the redback pulsar wind termination shock is much closer to the pulsar than that of pulsar wind nebulae, so we would
not expect the current sheet to fully dissipate the magnetic stripes \citep{2012SSRv..166..231R}. {We leave discussion of alternative particle acceleration scenarios to future studies.}

Outside the light cylinder, the magnetization in the pulsar wind should be (neglecting dissipation by the current sheet)
\begin{equation}
    \sigma_{pw}
    =\frac{B_{LC}^2}{4\pi \lambda n_{GJ}m_ec^2 },
\end{equation}
where $n_{GJ}$ is the Goldreich-Julian density and $\lambda$ is the multiplicity of $e^\pm$ pairs inside the magnetosphere \citep{2014ApJ...785L..33P, 2022ARA&A..60..495P}. Substituting $n_{GJ}=\Omega B_{LC}/2\pi c e$ for plasma co-rotating at the light cylinder \citep[e.g.][]{1969ApJ...157..869G}, we have
\begin{equation}
    \sigma_{pw}= \frac{\omega_B}{2\lambda \Omega}\approx 3\times10^5 \left(\frac{\lambda}{10^3}\right)^{-1}\left(\frac{P}{2 \text{ ms}}\right)^{-1}\left(\frac{B_{LC}}{10^5 \text{ G}}\right),
\end{equation}
where $\omega_B$ is the $e^\pm$ cyclotron frequency at the light cylinder. 
Our $\gamma_{max}$ suggest $\sigma\gtrsim10^5$ {for reconnection-powered acceleration}, so the pair multiplicity in these sources should be $\lambda\lesssim10^3$. This value is expected for MSPs and is lower than $\lambda \sim 10^5$ inferred for young pulsars with robust pair production \citep{2011ApJ...726L..10H, 2015ApJ...810..144T, 2022ARA&A..60..495P}. {Typical pair multiplicity values $\lambda= 400$ and $2000$ quoted in previous spider modeling \citep{2020ApJ...904...91V} are consistent with our limits. }

\bigskip
Our analysis of redback IBS X-ray spectra finds a synchrotron cooling break for J1723 at $E_{b,IBS}=7.4^{+1.3}_{-1.3}$\,keV, implying an IBS magnetic field of $38^{+3}_{-2}$\,G. This moderate local IBS magnetic field may be linked to the low viewing inclination of this source, where we might expect incomplete striped wind annihilation and a residual field in the post-shock flow at high latitudes. The presently deprecated J2215 break would imply a relatively high IBS field of $\sim70$\,G; if additional data require this break, $i$ may be at the low end of the fit range to accommodate incomplete striped wind annihilation. For J2339, we obtain upper limits on the field, consistent with its larger viewing angle near the spin equator.  The spectra also require maximum electron energy $\gamma_{max}>6\times10^4$ in all sources. The associated lower limits on $\sigma$ suggest a low pair multiplicity $\lambda\lesssim10^3$, supporting arguments that MSPs should have less efficient pair production than young pulsars. As sensitivities to hard X-ray spectral features and polarization improve, we can expect additional constraints on the particle acceleration mechanisms and magnetic fields in spider IBS shocks. 

\section*{Acknowledgments}
 The authors thank Alexander Philippov for useful discussions {and the anonymous referee whose comments improved our manuscript}. This work was supported in part by NASA grant 80NSSC23K1613. A.S. acknowledges the support of the Stanford University Physics Department Fellowship, the National Science Foundation Graduate Research Fellowship, and a Giddings Fellowship at the Kavli Institute for Particle Astrophysics and Cosmology at Stanford.

\bibliography{refs}{}
\bibliographystyle{aasjournalv7}



\end{document}